\documentclass[conference]{IEEEtran}

\ifCLASSINFOpdf
   \usepackage[pdftex]{graphicx}
\else
   \usepackage[dvips]{graphicx}
\fi

\usepackage[cmex10]{amsmath}
\usepackage{multicol}
\usepackage{comment}
\usepackage{amssymb}

\begin{document}

\title{Optimal County-Level Siting of Data Centers \\ in the United States}

\author{
\IEEEauthorblockN{Maria Vabson, Muhy Eddin Za'ter, Amir Sajadi, Kyri Baker, Bri-Mathias Hodge}
\IEEEauthorblockA{University of Colorado Boulder, Boulder, USA\\
\{maria.vabson, muhy.zater, amir.sajadi, kyri.baker, brimathias.hodge\}@colorado.edu}
}

\maketitle

\begin{abstract}

Data centers are growing rapidly, creating the pressing need for the development of critical infrastructure build out to support these resource-intensive large loads. Their immense consumption of electricity and, often, freshwater, continues to stress an already constrained and aging power grid and water resources. This paper presents a comprehensive modeling approach to determine the optimal locations to construct such facilities by quantifying their resource use and minimizing associated costs. The interdisciplinary modeling approach incorporates a number of factors including the power grid, telecommunications, climate, water use, and collocated generation potential. This work establishes the base model whose functionality is shown through several test cases focusing on carbon-free generation collocation on a county-level in the United States. The results suggest that while capital costs are the biggest driver, having a longer future outlook and allowing more variable generation collocation influences the model to choose sites with higher renewable potential. 

\end{abstract}

\begin{IEEEkeywords}
capacity expansion, data centers, optimization, power systems, renewable energy
\end{IEEEkeywords}


\section{Nomenclature}
\subsection{Indices}
\noindent
$i$: Collocated generation technologies \\ 
h: Time periods (hours) \\
l: Possible locations (county) \\
j: Network type $\{$transmission, telecommunications$\}$

\subsection{Parameters}
\noindent Y: Number of years \\
A: Aggregation value for to get to annual value for hourly data (ex. 91.25 for one representative seasonal day) \\
$\alpha^j$: Weight or inclusion of network j \\
$\rho$: Capital recovery factor \\
$\delta$: Discount or interest rate 

\subsection{Demand Parameters}
\noindent $C^D$: Data center total power capacity (MW) \\
$D^p_h$: Hourly data center energy demand (MWh) \\
$PUE_l$: Power usage effectiveness (kWh/kWh) \\
$WUE_l$: Water usage effectiveness (L/kWh) \\
$D^p_{h,l} = PUE * D^p$: Hourly data center energy demand (MWh) \\
$D^w_{h,l} = WUE * D^p$: Hourly data center water demand (L)

\subsection{Collocated Plant Parameters}
\noindent 
For i = $\{$wind, solar, geothermal$\}$: \\
$C^{max}_{i,l}$: Maximum possible capacity (MW) \\
$f_{i,h,l}$: Capacity factor, annual average as well as hourly values for wind and solar. \\
$P^{max}_{i,h,l} = f_{i,h,l}*C^{max}_{i,l}$: Maximum possible generation (MWh)\\
$I_i$: Investment capital cost for generation technology $(\$/MW)$\\
$V_{i,h}$: Variable O$\&$M cost of generation technology $(\$/MWh)$\\
$F_{i}$ : Fixed annual O$\&$M cost of generation technology $(\$/MWh-yr)$ \\
$\pi^{c}$: Curtailment Penalty ($\$/MWh$)\\
$\pi^{e}$: Price of selling back to grid ($\$/MWh$)\\
B: Renewable penetration fraction 

\subsection{Additional Collocated Plant Parameters} 
\noindent 
For i = $\{$nuclear SMR, gas, diesel, biomass, etc.$\}$, \\
$P^{max}_{i,h,l}$: Maximum possible generation (MWh)\\
$y_{i,h,l}$: Plant availability \\
$\pi^G_{op}$: Operating cost as fraction of CAPEX \\
$r_i$: Ramp rate per hour \\
$T_{i,h,l}$: Startup time required to start from cold start (hours) \\
$\pi^G_{start}$: Startup cost as a fraction of the hourly max output

\subsection{Storage Parameters}
\noindent 
$\eta$: Round trip efficiency \\
$S^{max}_{charge\_ rate}$: Maximum charging rate as fraction of $S_{cap}$ \\
$S^{max}_{discharge \_ rate} $: Maximum discharging rate as fraction of $S_{cap}$ \\
$S^{max}_{cap}$: Maximum storage capacity as fraction of $S_{cap}$ \\
$S^{min}_{cap}$: Minimum storage capacity as fraction of $S_{cap}$ \\
d: Self-discharge rate per hour \\
$\gamma^c$: Charging efficiency \\
$\gamma^d$: Discharging efficiency

\subsection{Network Parameters}
\noindent
$d^{j}_l$: Network line investment distance (miles) \\
$E^{j}_l$: Investment capital cost of network line $(\$/km)$ \\
$K_{l}(z_l)$: Cost of substation (new/upgrade) as a function of the transmission capacity for sufficient capacity\\
$\pi^{trans}_{h,l}$: Hourly price of electricity based on location $(\$/MWh)$ \\ 
H: Reserve margin headroom fraction

\subsection{Water Parameters}
\noindent
$\pi^{water}_{l}$: Hourly price of water use based on location $(\$/L)$ \\ 
$R_l$: Water scarcity risk based on location \\
$\pi^{r}$: Water scarcity risk penalty ($\$/L$)\\
$\alpha^w$: Weight or inclusion of network water parameters

\subsection{Decision Variables}
\[
x_{l} = \begin{cases}
    1, & \text{if site } l \text{ is selected}\\
    0, & \text{otherwise}
\end{cases}
\]
\noindent
$P^g_{h,l}$: The amount of energy from grid at location l, hour h (MW)\\
$P^{load}_{i,h,l}$: The amount of generated energy going to meet the data center load demand (MWh) \\
$P^{storage}_{i,h,l}$: The amount of generated energy going to charge storage (MWh) \\
$P^{e}_{i,h,l}$: The amount of generated energy being exported to the grid (MWh) \\
$P^{c}_{i,h,l}$: The amount of generated energy being curtailed (MWh) \\
$z_{l} \in \mathbb{Z}$ : Max transmission capacity (MW) \\
S$^{c}_{h,l}$: Amount of charge going storage  \\
S$^{d}_{h,l}$: Amount of charge from storage to load  \\
S$^{soc}_{h,l}$: State of charge of the storage \\ 

\section{Introduction}

Data centers are critical for the modern information-based society, and the digitization of the economy especially with the expansion of artificial intelligence (AI) use. Data centers accounted for about 1.5$\%$ of global electricity consumption in 2024 with projections estimating their electricity consumption to grow by about 15$\%$ per year between 2024 and 2030 \cite{ai2025iea}. In addition to the electricity consumption, data centers require fiber optic cables for telecommunications and in many cases freshwater to cool the facility. Fiber optics are essential to data center connectivity and their construction often comes with a hefty cost. For example, underground fiber optic deployment can cost anywhere between $\$40,000-\$75,000$ per kilometer \cite{fiber_development}. Similarly, data centers' operation can take a toll on the local water resources. It is projected that by 2027 the global AI water demand will surpass the annual water withdrawal of Denmark, or half of the UK, with around 5 billion cubic meters \cite{li2023making}. These demands are exacerbated in warmer and more humid climates that may require additional electricity and water for cooling \cite{turek2021optimized}.
Accordingly, the siting of new data centers should be wholistic and simultaneously consider all aspects of critical infrastructure and public utilities; namely, collocated generation and storage, the electric power grid, fiber optics, and water use. 

Previous studies on data center siting have highlighted utility prices, network infrastructure, and proximity to population centers as influencing siting decisions \cite{goiri2011}. For example, certain applications may strive for lower latency associated with closer proximity to users \cite{turek2021optimized}. In general, siting is a complex challenge, especially when considering renewable collocation. For example, the benefits of cooler climates in more northern locations coincides with lower potential for solar capacity \cite{depoorter2015location}. However, it has been found that collocating wind and solar with data centers can reduce costs without sacrificing reliability through optimized scheduling to maximize the use of renewable energy \cite{ huang2020review}. It has also been noted that one-fifth of data center servers are taking water from moderately to highly water-stressed watersheds, and nearly half of servers are at least partially powered by power plants located in water-stressed regions \cite{siddik2021environmental}. All of these factors, along with findings that even 10 km differences in location could account for approximately 9$\%$ energy savings \cite{turek2021optimized}, highlight the importance of wholistic siting processes. There is a body of literature that explores data center siting with its infrastructure considerations as demonstrated in Table \ref{tab:other_papers}. However, the power grid aspect of the existing literature either addresses the expansion of the transmission infrastructure or focuses on the potential of renewable energy, but not their simultaneous consideration and collocation. This paper aims to fill this gap by incorporating the data center capacity expansion problem with a consideration of collocated generation and storage. 

\begin{table*}[!htp]
\renewcommand{\arraystretch}{1.2}
\centering
\caption{Data Center Geographic Siting Parameter}
\label{tab:other_papers}
\begin{tabular}{c c c c c c c c c c c c c}
\hline
Work & Grid Expansion & On-Site Energy & Storage &  Fiber &  Water & Labor & Natural Hazard & Land & Climate & Granularity & Region\\
\hline
\cite{depoorter2015location} & & \checkmark &  & & & & & &   \checkmark & 5 Countries & Europe\\

\cite{arzumanyan2025geospatial} & \checkmark& &  & \checkmark & \checkmark & \checkmark & \checkmark & \checkmark & & Fine Scale &  Texas \\

\cite{ayyildiz2025location} & & \checkmark & & \checkmark & \checkmark & \checkmark & \checkmark & \checkmark & \checkmark & 6 Provinces & Turkey \\

\cite{guo2021integrated} & \checkmark & &      &\checkmark & \checkmark & & & \checkmark & \checkmark & 253 Regions & US \\

This work & \checkmark & \checkmark & \checkmark & \checkmark & \checkmark & &  & & \checkmark & 3066 Counties & US\\
\hline 
\end{tabular}
\end{table*}

This paper presents a comprehensive model for optimal data center siting in the United States with county-level granularity. This level of granularity is essential due to variability in climate, generation potential, and regional utility pricing. The model considers three main parameters that are critical for data center operation: electricity, water, and fiber optics. The electricity component of the model is a capacity expansion formulation that allows technology-agnostic collocated generation including non-renewable, renewable sources, and storage technologies. Also, the water needed for data center cooling is modeled along with the site-specific regional water scarcity risk, while the fiber optics build out is quantified by approximating the distance to existing fiber infrastructure. The overall model can serve as an adaptable tool for determining the best cost estimates of developing data centers. This model is formulated as a multi-objective optimization problem that evaluates factors concerning the construction and operations of data centers. The objective is to reduce the costs, negative impacts on surrounding communities, and broader environmental impacts of constructing data centers. Table \ref{tab:other_papers} compares the parameters and considerations of this paper with those of the most closely relevant papers in geospatially siting data centers. It highlights the core contributions of this paper, which are (1) simultaneously incorporating collocated generation and storage in the capacity and grid expansion model and (2) the granularity of the spatial resolution in siting analysis. 

As a case study, it evaluates data center siting at the county-level across the entire continental United States, which has not been previously explored. The use cases presented here highlight the functionality of possibly collocating data centers with carbon-free generation and storage. The findings of this paper serve to inform national decisions, with the ability to further incorporate the inclusion or exclusion of additional parameters in future work. 

\section{Model Formulation}

\begin{figure}[!htp]
    \centering
    \includegraphics[width=0.95\linewidth]{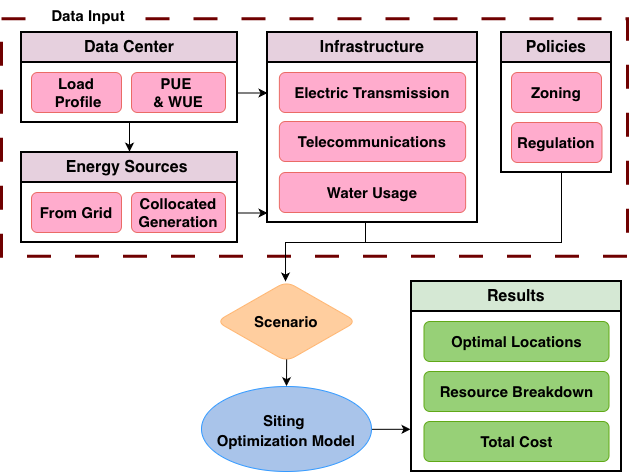}
    \caption{Model flow chart.}
    \label{fig:flowchart}
\end{figure}

Building off of a traditional power capacity expansion model, this model takes a wider perspective to include the system-wide impacts that large loads will create across different industries. It incorporates the amount of investment needed to construct collocated generation, build new transmission lines, substations, and fiber infrastructure. The capacity expansion optimization is based on minimizing the given capital, operational costs, and penalties of the electricity and water use. The optimization model structure is illustrated in Fig. \ref{fig:flowchart}. 

This nationwide county-level granularity provides a wide perspective on the possible future of data center siting. This model does not currently replace a site-specific feasibility study, in which it would be more pertinent to study the long term feasibility and challenges associated with specific locations. However, the model outputs provide insights to compare against current data center developments and the future prospective locations that may be targeted. The remainder of this section describes the notation and model formulation.

\subsection{Load Model}
The load input into the model is a seasonal profile that represents the demand variability and peak load times. This allows for the generalizability of this model to AI data centers, crypto currency mining facilities, and industrial loads. However, for the purpose of data centers specifically, a reference Information Technology (IT) load is input into the model and the total facility energy and water demand is determined by multiplying the given IT load by the Power Usage 
Effectiveness (PUE) and Water Usage Effectiveness (WUE) respectively \cite{shehabi20242024}. The energy and water use estimates are based off of the associated average annual PUE and WUE values of various facilities and the type of cooling technology used, based on the International Energy Conservation Code (IECC) climate regions \cite{climzonearticle}. The most common cooling technologies used for AI data centers include airside economizers with adiabatic cooling with either air or water cooled chillers, as well as IT Liquid cooling, also utilizing air or water cooled chillers \cite{shehabi20242024}. This model relies on PUE and WUE values for data centers using airside economizers and direct adiabatic cooling and using water-cooled chillers as supplemental cooling based on climate regions estimated by the work in \cite{lei2025water}. 

The main purpose of any capacity expansion model is to meet the load at each time step, which is represented by the constraint \eqref{eq:loadbalance}. This ensures there is sufficient generation with adequate headroom at every location and hour evaluated. The energy from the collocated generation, from potential storage discharge, and energy from the interconnected grid are the relevant decision variables reflecting the generation types that may be used to meet the load.
\begin{equation}
\sum_{i} P^{load}_{i,l,h} + P^g_{l,h} + S^{d}_{h,l} \ge R * D^p_h \ \ \ \ \forall h,l
\label{eq:loadbalance}
\end{equation} 

Once the data center load is established, the main question of where to site such a data center can then be answered. Based on the size of the load and its profile, the types of generation and their respective costs are then minimized in the objective function, subject to additional operational constraints. 
\subsection{Objective Function}
The objective function \eqref{eq:objective} minimizes the sum of capital costs, operational costs, and penalties. In order to account for different requirements and their relative importance when siting a data center, the objective function is formulated with modularity in place. This design allows for flexible customization of the objective and priorities with the ability to add coefficients to enhance or diminish the influence of certain considerations. The lifetime (Y), capital recovery factor ($\rho$), and coefficients ($\alpha$) in the subsections of the objective function provide flexibility into the modeling of specific components, and capital and operational costs.

\begin{equation}
\begin{split}
    {\min} \sum_{l}\sum_{i} x_l \Bigl( \rho C^{max}_{i,l} I_i + YC^{max}_{i,l} F_i \Bigl) \\
     + \sum_{l}x_l \Bigl( \rho \Bigr[ \sum_{j}(\alpha^jE^j_l z^j_l )+ K_{l} \Bigr] + \alpha^w R_l \pi^{w}_{risk} \Bigl) \\
    + \sum_{l}\sum_{i}\sum_{h} x_l Y A \Bigr[ P^{load}_{i,l,h} V_{i,h} + P^{c}_{i,h,l}\pi^{c} + \alpha^w D^w_h\pi^{w}_{h,l}\\ 
    + \alpha^{trans} \bigl( P^g_{l,h} \pi^{trans}_{h,l} - P^{e}_{i,h,l} \pi^{e} \bigl) \Bigr]
\end{split}
\label{eq:objective}
\end{equation} 

The summation in the first line adds together the capital and fixed costs of new generators and storage technologies. The second summation adds up the cost of building transmission, telecommunication, and water risk if specified to be included in the model by the respective $\alpha$ parameter. The final summation takes all of the variable operational costs of the energy generation, energy from the grid, water use, costs of curtailment, and costs or income of the excess collocated energy produced. Every part of the objective function is multiplied by the primary decision variable $x_l$ determining if a data center will be placed at that location. It is subject to the number of locations of interest to be selected as restricted by \eqref{eq:minloc} and the rest of the constraints that follow. 
\begin{equation}
\text{minimum } \# \text{ of locations} < \sum_{l}x_l
\label{eq:minloc}
\end{equation}
The capital and fixed costs for the collocated generation and storage are summed in the first line subject to the relevant capital recovery factor calculated based on a set discount or interest rate that relates the capital to operational costs for a more representative present value consideration \eqref{eq:crf} \cite{crfnreal}. 
\begin{equation}
\rho = (\delta (1 + \delta )^Y) / ([(1 + \delta )^Y]-1)
\label{eq:crf}
\end{equation}
The modularity of this framework takes into consideration different user specifications subject to the $\alpha$ coefficients that can be set to include or weigh the specific factors, such as additional build out of the transmission grid, telecommunications, or communications infrastructure.

\subsection{Collocated Generation Model}
The non-renewable or fuel-based generation sources are modeled by several different aspects such as their minimum and maximum output \eqref{eq:minoutput}, the maximum ramp up and down rates \eqref{eq:rampup}, \eqref{eq:rampdown}, and the startup time \eqref{eq:startup}. The following constraints are applied to nuclear small modular reactors (SMR), diesel, biomass, and natural gas generators. They are not applied to i = $\{$solar, wind, geothermal$\}$. 

\begin{equation}
    P^{min}_{i,l} * y_{i,h,l} \le P^{load}_{i, h,l} \le P^{max}_{i,l} * y_{i,h,l}
\label{eq:minoutput}
\end{equation}
\begin{equation}
     P^{load}_{i, h,l} - P^{load}_{i, h-1,l}  \le r_{i,l}
\label{eq:rampup}
\end{equation}
\begin{equation}
     P^{load}_{i,h-1,l} - P^{load}_{i, h,l}  \le r_{i,l}
\label{eq:rampdown}
\end{equation}
\begin{equation}
T_{i,h,l} \ge
\begin{cases}
y_{i,h,l}, & \text{if } h = 0, \\[6pt]
y_{i,h,l} - y_{i,h-1,l}, & \text{if } h > 0.
\end{cases}
\label{eq:startup}
\end{equation}

The startup time \eqref{eq:startup} is factored into the variable operational costs by multiplying the startup time by the startup costs of that plant every time it needs to be restarted.

\subsection{Renewable Resource Model}
For i = $\{$wind, solar$\}$, the variable generation (VG) is modeled to account for the power that supplies the load in addition to charging the storage, and being curtailed or exported since these are not dispatchable sources of energy. So, the total amount of possible energy generation is the maximum limit of the sum of the energy used to meet the data center load, charge the storage, be curtailed or be exported to the grid seen as seen in \eqref{eq:rengen}. The maximum possible generation at a location at a given hour is found by multiplying the capacity of that location by the associated average annual capacity factor or hourly value if available. 
\begin{equation}
     P^{load}_{i,h,l} + P^{storage}_{i,h,l} + P^{e}_{i,h,l} + P^{c}_{i,h,l} \leq P^{max}_{i,h,l} 
\label{eq:rengen}
\end{equation}

Due to stability concerns of using 100$\%$ VG, the following constraint \eqref{eq:renpen} can be used to limit the amount of generation from solar and wind being used to power the data center directly.
\begin{equation}
    \sum_{i}P^{load}_{i,h,l} \le D^p_h * B 
\label{eq:renpen}
\end{equation}

Geothermal is modeled as a non-VG source since it is dispatchable. Thus, it does not contribute to storage, export, and should not be curtailed. So, for i = {geothermal} the generation is modeled by \eqref{eq:geogen}.
\begin{equation}
     P^{load}_{i,h,l} \leq P^{max}_{i,h,l} 
\label{eq:geogen}
\end{equation}

\subsection{Storage Model}
For this setup, only the VG is able to charge the storage \eqref{eq:rentocharge}. 
For i = $\{$wind, solar$\}$,
\begin{equation}
   \sum_i P^{storage}_{i,h,l} = S^{c}_{h,l}
\label{eq:rentocharge}
\end{equation}
To reiterate, the reason for exclusion of geothermal here, even though it is renewable, it can be dispatched.
The rate of charging and discharging is limited the maximum rates \eqref{eq:maxcharge} and \eqref{eq:maxdishcharge} respectively, with the maximum discharge which is powering the load must be below the load within a margin \eqref{eq:storagetoload}.
\begin{equation}
   S^{c}_{h,l} \le S^{max}_{charge\_ rate}
\label{eq:maxcharge}
\end{equation}
\begin{equation}
   S^{d}_{h,l} \le S^{max}_{discharge\_ rate}
\label{eq:maxdishcharge}
\end{equation}
\begin{equation}
    S^{d}_{h,l} \le D^p_h * B 
\label{eq:storagetoload}
\end{equation}

To maintain and keep track of the state of charge (SOC), the model starts with the SOC at half of the storage capacity for each location for simplicity and subsequently is calculated by the amount of charging and discharging occurring \eqref{eq:soc}, subject to the maximum and minimum limits \eqref{eq:soclimits}.
\begin{equation}
S^{soc}_{h,l} =
\begin{cases}
0.5 *S^{max}_{cap}, & \text{if } h = 0, \\[6pt]
(1-d)S^{soc}_{h-1,l} + \gamma^c S^{c}_{h,l} - \gamma^d S^{d}_{h,l} , & \text{if } h > 0.
\end{cases}
\label{eq:soc}
\end{equation}
\begin{equation}
   S^{min}_{} \le  S^{soc}_{h,l} \le S^{max}_{cap}
\label{eq:soclimits}
\end{equation}

\subsection{Network Model}
To determine the transmission rating necessary for a data center to connect to a power grid, the rating is set to be larger than the total projected electric power exchange with the grid with some extra capacity as contingency capacity using constraint \eqref{eq:gridcapacity}. The substation capacity is then set to the same rating as the transmission. 
\begin{equation}
    H *(P^g_{h,l} + \sum_{i}P^{e}_{i,h,l}) \le z_l ~~~\text{for } i = \{\text{wind, solar}\}
\label{eq:gridcapacity}
\end{equation}

For the telecommunications aspect, the amount of fiber build-out necessary is approximated based on the aggregation of the major US cities that have fiber optic connectivity \cite{durairajan2015intertubes}. The cost of building the fiber is calculated by multiplying the cost of deployment per unit length by the distance from the siting locations of interest to the city nodes \cite{fiber_development}.


\section{Results and Discussion}

To demonstrate the application of this capacity expansion data center siting model, two scenarios were analyzed. 
The first case, labeled Base Case, represents a data center siting problem considering collocated solar, wind, geothermal ($4500$ m), nuclear SMR, and all contributing factors evaluated: transmission system expansion, fiber optics, and water usage. The second category of scenarios is built upon the Base Case to represent the conditions needed to power a data center fully with VGs (solar and wind only) and storage by setting different limits on their ability to supply the load.

For all cases, the Gurobi solver \cite{gurobi} was used to solve the optimization model and minimize the objective function \eqref{eq:objective} modeled using Pyomo in Python \cite{hart2011pyomo, bynum2021pyomo}. All data used for this analysis is open source except for the telecommunications nodes from \cite{durairajan2015intertubes}. The entire data processing, sources, and siting model is at this link\footnote{https://github.com/HodgeLab/datacenter-siting-model}.

\subsection{Base Case}
The Base Case represents the ability to add generation, in this case low carbon options, water use, as well as the build out of additional transmission, substations, and fiber optics. The values listed in Table \ref{tab:ren_pen_scenarios} were used here as a representative price of exporting electricity and penalizing curtailment. Moreover, this analysis assumed the use of lithium-ion battery storage through the adoption of its parameters' values, the commercial availability of nuclear SMRs, a defined singular data center large load profile, adiabatic cooling, and seasonal aggregation of hourly daily load variations to be able to account for storage and ramping details.
\begin{table}[!ht]
\renewcommand{\arraystretch}{1.3}
\centering
\caption{Modeling Assumptions}
\label{tab:ren_pen_scenarios}
\begin{tabular}{c c}
\hline
Parameter & Value \\
\hline
Data center Size         &  200 MW \\
Storage: Lithium Ion Battery & 150 MW\\
Export Price: $\pi^e$       &  5 MWh \\
Curtailment Penalty: $\pi^c$  &  20 $\$/MWh$\\
Years: Y                     & 20\\
Discount Rate & 1.2$\%$\\
\hline
\end{tabular}
\end{table}
Only low carbon collocated energy sources were modeled meaning reduced on-site greenhouse gas, not accounting for the power sources from the grid. This implies the potential for low carbon collocated generation sources such as solar, wind, geothermal, nuclear SMR, and battery storage for data center expansion and sustainable siting considerations along with fiber optic expansion and water use. The goal was to determine the possibility of fully powering next generation data centers with only low carbon energy sources. 

To this end, the solar, wind, and geothermal data was aggregated to select the location within a given county with the highest total renewable capacity available as an upper limit on the amount of renewable generation, with capacity determined from the NREL supply curves \cite{oedi_6001, oedi_6119, lopez2025renewable}. Hourly capacity factors for the wind and solar generation estimates were acquired from the NREL Regional Energy Deployment System (ReEDS) model \cite{hourlycfdata}. Location restrictions followed the land exclusion scenarios of the solar and wind supply curves, focused on the least restrictive open access scenario to get a better sense of total possible potential nationwide. This was the least limiting spatial restriction that still accounted for a number of protected lands, conservation zones, and areas used for the Department of Defense \cite{lopez2024solar}. 
In addition, to minimize computation, the hourly profiles were averaged to four seasonal days, those $96$ hours representing a one year profile. While this does not represent extreme weather events, this was a base estimate to help reflect the typical cooling needs, the amount of solar and wind capacity which varies hourly and seasonally, and the feasibility with respect to generators' operational constraints. 

\begin{figure}[h]
    \centering
    \includegraphics[width=1\linewidth]{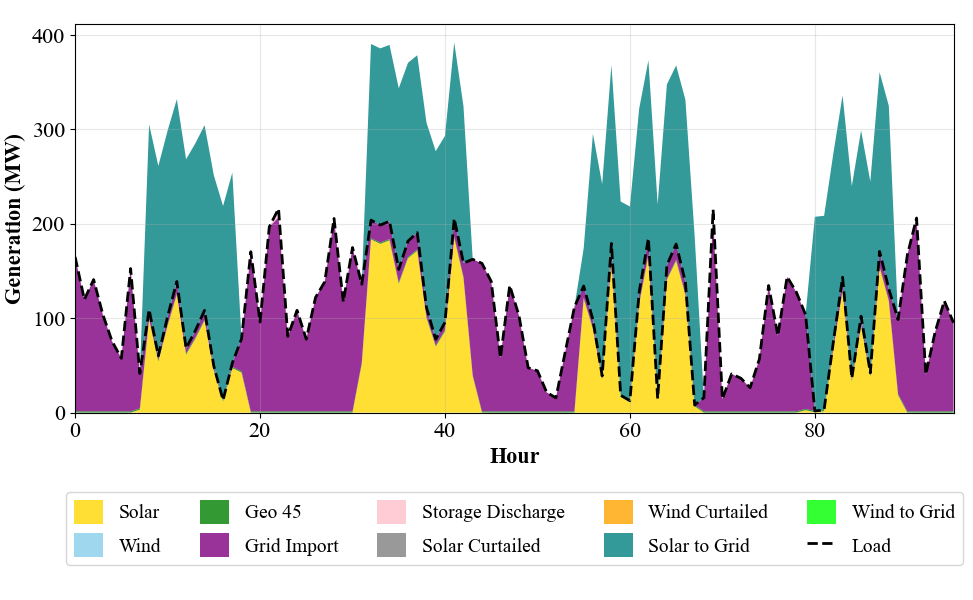}
    \caption{Hourly Generation Dispatch for Base Case}
    \label{fig:hourlygen}
\end{figure}

\begin{figure*}[!]
    \centering
    \includegraphics[width=0.8\linewidth]{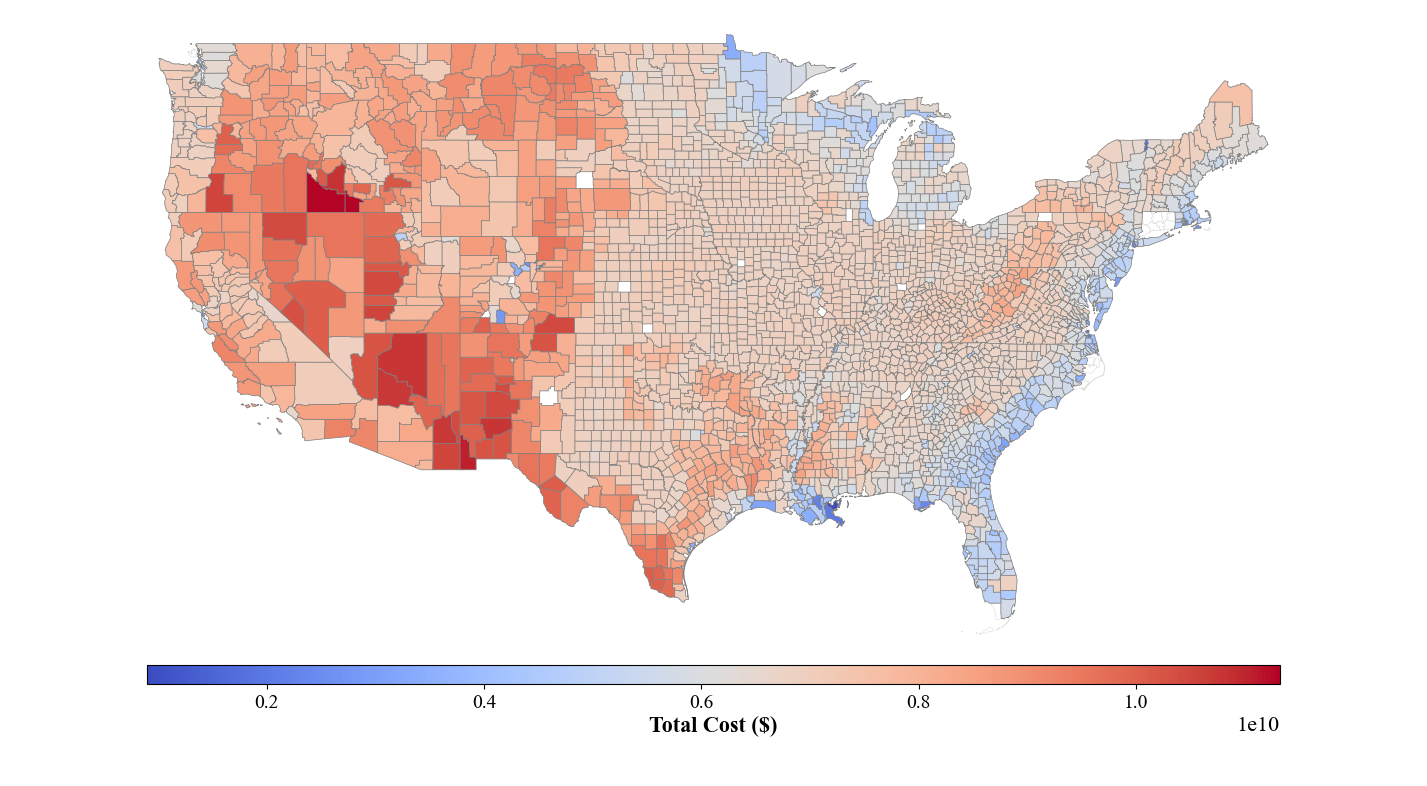}
    \caption{Heat Map Based on Optimization Objective for Base Case.}
    \label{fig:countybasecase}
\end{figure*}

The results of this analysis are shown in Fig. \ref{fig:hourlygen}. The example data center load profile is distinguished by the black dotted line and the generation output from the model used to meet the demand for the lowest cost location are in color. This load profile was randomly synthesized to reflect a more variable scenario, exaggerating the inference processing of an AI data center, compared to the more stable and constant energy usage during the training phase of AI. The outcome prioritizes solar and grid imports. This is likely due to the benefit of installing less new capacity while the interconnection to the electric grid allows for excess solar energy to be exported, as shown in teal. 

\begin{figure}[h]
    \centering
    \includegraphics[width=1\linewidth]{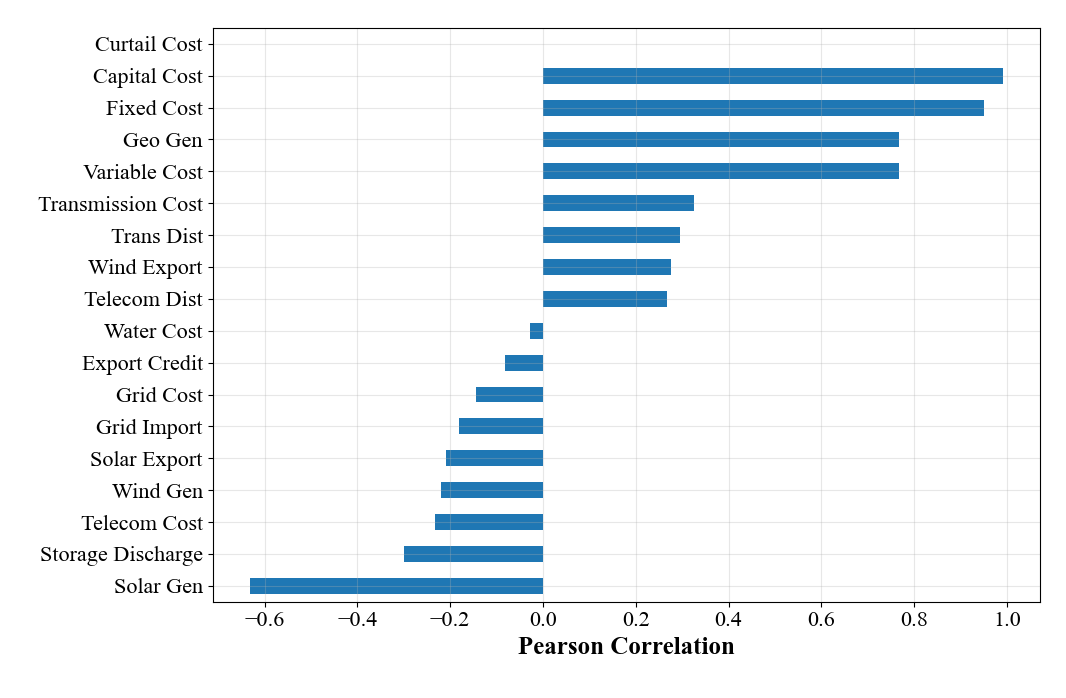}
    \caption{Pearson Correlation for Base Case.}
    \label{fig:countybasecasepearson}
\end{figure}

As a central result of this analysis, a county-level heatmap of the total cost calculated for the same scenario was generated and is displayed in Fig. \ref{fig:countybasecase}. The most expensive counties are represented in red and the least expensive in blue. The overall trend suggests lower costs occurred in the southern and eastern regions of the US. With the typical generation breakdown seen in Fig. \ref{fig:hourlygen}, this may be clear with the higher solar capacity factors in those regions being able to power the potential data centers. To understand the driving factors behind the selection of the locations at hand, a sensitivity analysis of the parameters was then conducted whose results are shown in Fig. \ref{fig:countybasecasepearson}. The Pearson correlation is a calculation of the linear relationship between the parameter and its contribution to the total objective function value. Applying this correlation here uncovered the driving factors behind the site selection of a data center as illustrated in Fig. \ref{fig:countybasecasepearson}. Another interesting insight is that geothermal generation is highly correlated with higher costs, suggesting that if geothermal generation was commercially available in a location, it would increase costs due to its larger investment and operating costs compared to solar and wind.
\subsection{Collocated Renewable Scenarios}
\begin{figure}[h]
    \centering
    \includegraphics[width=1\linewidth]{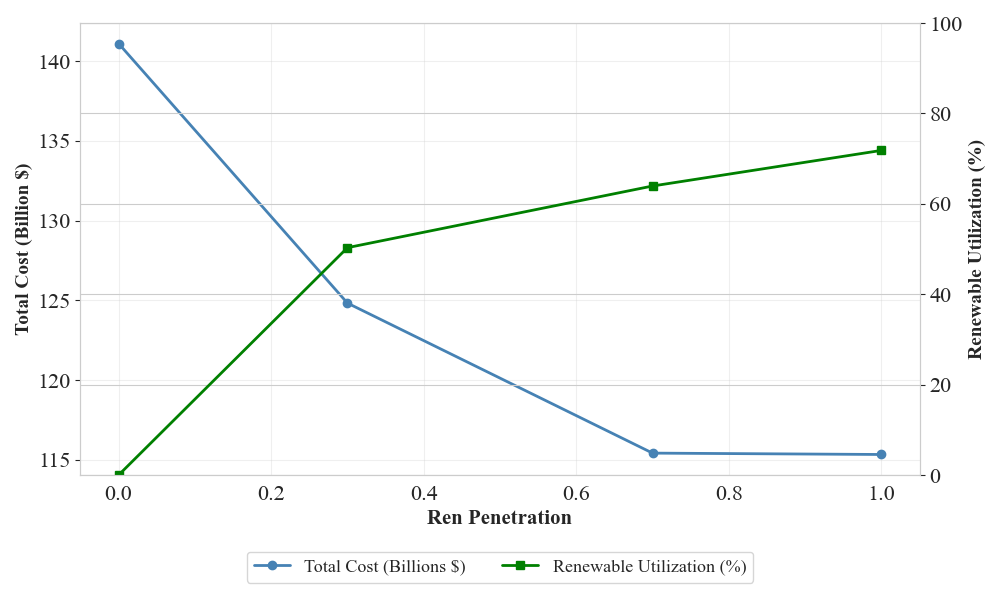}
    \caption{The overall cost depending on allowable VG collocation (left, blue). The amount of VG utilization related to the amount of allowable VG collocation (right, green).}
    \label{fig:renpentrationlines}
\end{figure}

In the subsequent analysis, the level of collocated VGs supporting the data center load was varied to determine the impact on siting and cost. This understanding is critical in decision-making and system planning, especially considering the ongoing interest of grid operators and regulators in adoption of VGs. While the amount of VG continues to grow, there are concerns about the impact on stability and the fact that they are not dispatchable. However, if possible to use them to a larger extent, especially with the storage capacities available, it seems like a feasible option to power future data centers. For this evaluation, increasing levels of penetration were tested and the subsequent impact on total cost over the set project lifetime along with the amount of VG used are shown in Fig. \ref{fig:renpentrationlines}. They show that the use of VGs monotonically rises (green trace) as the amount allowed was increased which also coincided with lower overall costs proportional to increase in the collocated VG use (blue trace).
Looking at these results, it can be inferred that allowing more collocated VGs to power the data centers could be beneficial to lowering operational costs whilst also promoting carbon free energy and indirectly reducing greenhouse gas emissions. 

The model developed in this paper could serve as a tool for identifying and comparing suitable locations for new data centers. It is designed to be modular, scalable, and customizable to projects' specified inputs, priorities, and requirements. This approach allows more flexibility and comprehensive planning capability. This is a timely and important development as the current trends prioritize locations with cheaper electricity, water, and land costs, looking at the near-term costs. However, based on the presented results, it is recommended to prioritize investments in collocated VGs to power data centers, noting the total costs proportionally decline in the long-term with more use of collocated VGs. In addition, the regions currently with lower costs are often in warmer or more humid climates which will continue to put more strain on the electric grid due to higher cooling needs and the lower transmission ratings due to line thermal limits, and may not be sustainable in terms of continued development. The test cases examined here showed that there is a potential for being able to site data centers in a manner that incorporates carbon-free generation and could help minimize costs across the board. 

\section{Conclusion}

This paper developed a model for data center siting with county-level granularity. It considered the key factors that influence cost and decision making for their siting, while also shedding light on possible considerations that would guide choices for more sustainable development. Due to the flexibility of the modeling framework, it is applicable for the siting of any large industrial facilities with significant power and water dependencies. The results highlight that more use of collocated VGs in powering data centers will help reduce their overall cost of operation. Further work will interoperate other factors to make this model more comprehensive. These factors could include the cost of the land, labor cost, and natural hazard. 

\section{Acknowledgement}
The work has been supported by the Climate Innovation Collaboratory (CIC), an ongoing alliance between Deloitte and the University of Colorado Boulder.


\bibliographystyle{ieeetr}
\bibliography{sources}

@techreport{lopez2024solar,
  title={Solar Photovoltaics and Land-Based Wind Technical Potential and Supply Curves for the Contiguous United States (2023 Edition)},
  author={Lopez, Anthony and Pinchuk, Pavlo and Gleason, Michael and Cole, Wesley and Mai, Trieu and Williams, Travis and Roberts, Owen and Rivers, Marie and Bannister, Mike and Thomson, Sophie-Min and others},
  year={2024},
  institution={National Renewable Energy Laboratory (NREL)},
  url = {https://doi.org/10.2172/2283517}
}

@article{siddik2021environmental,
  title={The environmental footprint of data centers in the United States},
  author={Siddik, Md Abu Bakar and Shehabi, Arman and Marston, Landon},
  journal={Environmental Research Letters},
  year={2021},
  publisher={IOP Publishing}}

@techreport{shehabi20242024,
    author = {Shehabi, Arman and Hubbard,  Alex and Newkirk, Alex and Lei, Nuoa and Siddik, Md Abu Bakkar and Holecek, Billie and Koomey, Jonathan and Masanet, Eric and Sartor, Dale and others},
    title = {2024 United States Data Center Energy Usage Report},
    institution = {Lawrence Berkeley National Laboratory},
    year = {2024}
}

@article{turek2021optimized,
  title={Optimized data center site selection—Mesoclimatic effects on data center energy consumption and costs},
  author={Turek, Dirk and Radgen, Peter},
  journal={Energy Efficiency},
  volume={14},
  number={3},
  pages={33},
  year={2021},
  publisher={Springer}}

@article{huang2020review,
  title={A review of data centers as prosumers in district energy systems: Renewable energy integration and waste heat reuse for district heating},
  author={Huang, Pei and Copertaro, Benedetta and Zhang, Xingxing and Shen, Jingchun and L{\"o}fgren, Isabelle and R{\"o}nnelid, Mats and Fahlen, Jan and Andersson, Dan and Svanfeldt, Mikael},
  journal={Applied energy},
  volume={258},
  pages={114109},
  year={2020},
  publisher={Elsevier}}

@article{li2023making,
  title={Making ai less" thirsty": Uncovering and addressing the secret water footprint of ai models},
  author={Li, Pengfei and Yang, Jianyi and Islam, Mohammad A and Ren, Shaolei},
  journal={Communications of the ACM},
  Volume = {68},
  Issue = {7},
  pages = {54--61},
  year={2023}}

@inproceedings{goiri2011,
  title={Intelligent placement of datacenters for internet services},
  author={Goiri, {\'I}nigo and Le, Kien and Guitart, Jordi and Torres, Jordi and Bianchini, Ricardo},
  booktitle={2011 31st International Conference on Distributed Computing Systems},
  pages={131--142},
  year={2011},
  organization={IEEE}
}

@misc{climzonearticle,
    author = {Baechler, Michael and Gilbride, Theresa and Cole, Pam  and Hefty, Marye and Ruiz, Kathi},
    title = {Building America Best Practices, Volume 7.3 - Guide to Determining Climate Regions by County},
    institution = {Pacific Northwest National Laboratory},
    year = {2015},
    url = {https://www.energy.gov/eere/buildings/articles/building-america-best-practices-series-volume-73-guide-determining-climate}
}

@article{ayyildiz2025location,
  title={Location selection methodology for data center with renewable energy integration},
  author={Ayyildiz, Ertugrul and Yildirim, Betul and Aydin, Nezir},
  journal={Renewable Energy},
  pages={123270},
  year={2025},
  publisher={Elsevier}
}

@article{lei2025water,
  title={The water use of data center workloads: A review and assessment of key determinants},
  author={Lei, Nuoa and Lu, Jun and Shehabi, Arman and Masanet, Eric},
  journal={Resources, Conservation and Recycling},
  volume={219},
  pages={108310},
  year={2025},
  publisher={Elsevier}}

@article{depoorter2015location,
  title={The location as an energy efficiency and renewable energy supply measure for data centres in Europe},
  author={Depoorter, Victor and Or{\'o}, Eduard and Salom, Jaume},
  journal={Applied Energy},
  volume={140},
  pages={338--349},
  year={2015},
  publisher={Elsevier}
}

@techreport{ai2025iea,
    author = {IEA},
    title = {Energy and AI},
    institution = {IEA},
    year = {2025}
}

@article{arzumanyan2025geospatial,
  title={Geospatial Suitability Analysis for Data Center Placement: A Case Study in Texas, USA},
  author={Arzumanyan, Mariam and Rodriguez Calzado, Edna and Lin, Ning and Bahadur, Vaibhav and Dasx, Jani and Ko, Lucy Tingwei and Koesterke, Lars},
  journal={Sustainable Cities and Society},
  year={2025}
}

@article{guo2021integrated,
  title={Integrated energy systems of data centers and smart grids: State-of-the-art and future opportunities},
  author={Guo, Caishan and Luo, Fengji and Cai, Zexiang and Dong, Zhao Yang},
  journal={Applied Energy},
  volume={301},
  pages={117474},
  year={2021},
  publisher={Elsevier}}

@misc{gurobi,
  author = {{Gurobi Optimization, LLC}},
  title = {{Gurobi Optimizer Reference Manual}},
  year = 2024,
  url = "https://www.gurobi.com"
}

@misc{crfnreal,
    author = {NREL},
    title = {Simple Levelized Cost of Energy (LCOE) Calculator Documentation},
    year = {2025},
    url = {https://www.nrel.gov/analysis/tech-lcoe-documentation}
}

@book{bynum2021pyomo, title={Pyomo--optimization modeling in python}, author={Bynum, Michael L. and Hackebeil, Gabriel A. and Hart, William E. and Laird, Carl D. and Nicholson, Bethany L. and Siirola, John D. and Watson, Jean-Paul and Woodruff, David L.}, edition={Third}, volume={67}, year={2021}, publisher={Springer Science \& Business Media} }

@article{hart2011pyomo, title={Pyomo: modeling and solving mathematical programs in Python}, author={Hart, William E and Watson, Jean-Paul and Woodruff, David L}, journal={Mathematical Programming Computation}, volume={3}, number={3}, pages={219--260}, year={2011}, publisher={Springer} }

@misc{fiber_development,
title = {Fiber Deployment Annual Report},
institution = {Fiber Broadband Association},
url = {https://fiberbroadband.org/wp-content/uploads/2024/01/Fiber-Deployment-Annual-Report-2023_FBA-and-Cartesian.pdf},
year = {2023},
}

@inproceedings{durairajan2015intertubes,
  title={InterTubes: A study of the US long-haul fiber-optic infrastructure},
  author={Durairajan, Ramakrishnan and Barford, Paul and Sommers, Joel and Willinger, Walter},
  booktitle={Proceedings of the 2015 ACM Conference on Special Interest Group on Data Communication},
  pages={565--578},
  year={2015}}

@misc{oedi_6001, title = {United States Utility-Scale PV Supply Curves 2023},
author = {NREL},
url = {https://data.openei.org/submissions/6001},
place = {United States}, year = {2023}, month = {06}}

@misc{oedi_6119, title = {United States Land-based Wind Supply Curves 2023}, 
author = {NREL}, 
url = {https://data.openei.org/submissions/6119}, journal = {}, 
place = {United States}, year = {2023}, month = {06}}

@techreport{lopez2025renewable,
  title={Renewable Energy Technical Potential and Supply Curves for the Contiguous United States: 2024 Edition},
  author={Lopez, Anthony and Zuckerman, Gabriel R and Pinchuk, Pavlo and Gleason, Michael and Rivers, Marie and Roberts, Owen and Williams, Travis and Heimiller, Donna and Thomson, Sophie-Min and Mai, Trieu and others},
  year={2025},
  institution={NREL},
  url = {https://docs.nrel.gov/docs/fy25osti/91900.pdf}
}

@misc{hourlycfdata,
  title   = {2024 County-Level Hourly Renewable Capacity Factor Dataset for the ReEDS Model},
  year    = {2024},
  author = {NREL},
  url     = {https://catalog.data.gov/dataset/2024-county-level-hourly-renewable-capacity-factor-dataset-for-the-reeds-model}
}

\end{document}